\documentclass[12pt]{article}
\usepackage{amsmath,amssymb,amsfonts, epsfig}
\usepackage{color}
\usepackage{bm}

\makeatletter
\@addtoreset{equation}{section}
\makeatother

\begin{document}
\title{
\begin{flushright}
\ \\*[-80pt] 
\begin{minipage}{0.2\linewidth}
\normalsize
KUNS-XXXX  \\*[50pt]
\end{minipage}
\end{flushright}
{\Large \bf 
A model of quarks with $\Delta (6N^2)$ family symmetry
\\*[20pt]}}

\author{
Hajime~Ishimori, $^{1}$ \
Stephen~F.~King,$^{2}$ \
\\*[20pt]
\centerline{
\begin{minipage}{\linewidth}
\begin{center}
$^1${\it \normalsize 
Department of Physics, Kyoto University, 
Kyoto 606-8502, Japan} \\
$^2${\it \normalsize
School of Physics and Astronomy,
University of Southampton,
Southampton, SO17 1BJ, U.K.}\\
\end{center}
\end{minipage}}
\\*[50pt]}
\vskip 2 cm
\date{\small
\centerline{ \bf Abstract}
\begin{minipage}{0.9\linewidth}
\medskip 
We propose a first model of quarks based on the discrete
family symmetry $\Delta (6N^2)$
in which the Cabibbo angle is correctly determined
by a residual $Z_2\times Z_2$ subgroup,
and the smaller quark mixing angles may be qualitatively understood from the model.
The present model of quarks may be regarded as a first 
step towards formulating a complete model of quarks and leptons based on $\Delta (6N^2)$,
in which the lepton mixing matrix is fully determined by a Klein subgroup.
For example, the choice $N=28$ provides an accurate determination of both the reactor angle
and the Cabibbo angle.
\end{minipage}
}

\begin{titlepage}
\maketitle
\thispagestyle{empty}
\end{titlepage}

\section{Introduction}

Neutrino oscillation experiments have discovered
large solar and atmospheric mixing angles in the lepton sector, together with a Cabibbo-sized reactor angle
$\theta^\ell_{13}$~\cite{pdg}.
In the approximation with $\theta^\ell_{13} \approx 0$, 
the tribimaximal mixing matrix is a quite interesting 
ansatz for the lepton sector~\cite{Harrison:2002er}.
The tribimaximal mixing ansatz led to 
a number of studies based on non-Abelian discrete flavor symmetries 
(see for review Refs.~\cite{Altarelli:2010gt,Ishimori:2010au,King:2013eh,King:2014nza}.)
In the direct approach, first a non-Abelian flavor symmetry $G^{(\ell)}_f$ 
for the lepton sector is assumed.
Then, such a symmetry is broken to $G_\ell$ ($G_\nu$) in the mass terms 
of the charged lepton (neutrino) sector.
It was also found that certain preserved subgroups of small discrete family symmetry
groups such as $S_4=\Delta (24)$, namely 
 $G_\nu = Z_2 \times Z_2$ and  $G_\ell=Z_3$,  
lead to simple mixing patterns such as tri-bimaximal mixing
matrix~\cite{resdS4}.
Recent neutrino experiments show that $\theta^\ell_{13} \neq 0$~\cite{rct,acl}.
However, the above direct approach 
is still interesting to derive experimental values of lepton mixing angles  
although we need much larger groups than $S_4 =\Delta (24)$
\cite{King:2013eh,King:2014nza}, for example $\Delta (6N^2)$ 
for large $N$ values such as $N=28$ \cite{King:2013vna}.

Here we consider such a direct approach applied to the quark sector in order to predict the CKM matrix.
Just as in the charged lepton sector where the residual symmetry 
$G_\ell$ may be in general $Z_l$~\cite{Lam:2007qc}, so in the quark sector one may envisage a residual
$Z_n\times Z_m$ symmetry of the quark mass matrices,
where this is a subgroup of some family symmetry.
However, in the quark sector, this approach is more 
challenging since larger mixing angles follow more directly from discrete family symmetry
than the small mixing angles present in the quark sector. 
Nevertheless the Cabibbo angle $\theta_{C}\approx \pi/14\approx 0.22$
has been shown to emerge from a residual $Z_2\times Z_2$ symmetry,
arising as a subgroup of the dihedral family symmetry 
$D_{7}$ \cite{Lam:2007qc,Blum:2007jz}, 
$D_{12}$ \cite{Kim:2010zub}, or $D_{14}$ \cite{Blum:2009nh,Blum:2007nt,Hagedorn:2012pg}. 
A more general analysis based on larger discrete family symmetry groups
was considered by \cite{Holthausen:2013vba,Araki:2013rkf}. 
Some analyses have considered both the lepton mixing angles and the Cabibbo angle as
arising from the same discrete family symmetry group
\cite{Hagedorn:2012pg,Holthausen:2013vba,Araki:2013rkf}. 
In all these works, only the Cabibbo angle is determined, since the residual $Z_2\times Z_2$ symmetry
only fixes the upper $2\times 2$ block of the mixing matrix. 
The other angles will appear by introducing small breaking terms 
for the symmetry at the next-to-leading order. 
A complementary approach to deriving the Cabibbo angle of $\theta_{C}\approx 1/4 $
at leading order was recently considered in an indirect model
based on a vacuum alignment $(1,4,2)$ without any residual symmetry
\cite{King:2013hoa}, although we shall not pursue such an indirect approach here.

In the present paper, we shall propose a model of quarks based on the discrete
family symmetry $\Delta (6N^2)$, following the above the direct approach to predicting the 
Cabibbo angle. This is the first model of quarks in the literature based on the $\Delta (6N^2)$ series.
Unlike the dihedral groups, $\Delta (6N^2)$ contains triplet representations
and is capable of fixing all the lepton mixing angles using the direct approach based on 
the full Klein symmetry subgroup preserved in the neutrino sector, where $N=28$ for example
gives both an accurate determination of the reactor angle \cite{King:2013vna}
and the Cabibbo angle \cite{Araki:2013rkf}.
Therefore the present model of quarks may be regarded as a first 
step towards formulating a complete model of quarks and leptons based on $\Delta (6N^2)$.
As above, we assume the residual symmetry for the quark sector 
to be a simple $Z_2\times Z_2$ symmetry corresponding to a $Z_2$ symmetry in each of the up and down sectors,
where the $Z_2$ symmetries are subgroups of a family symmetry $\Delta (6N^2)$.
Since the eigenvalues of $Z_2$ is $\pm 1$, 
at least two eigenvalues in $3\times 3$ matrices should be the same. 
With the phase difference of $\theta_{12}$ for up and down quark sectors, 
the Cabibbo angle is predicted by $\theta_C=\pi n/N$ where 
$n$ and $N$ are integers relating to the flavor symmetry. 

The motivation for constructing an explicit model of quarks in this approach, is that the 
$Z_2\times Z_2$ symmetry only determines the Cabibbo angle, and a concrete model
is required in order to shed light on the remaining small quark mixing angles
$\theta_{23}$ and $\theta_{13}$ which are not fixed by the symmetry alone.
Within the specified model,
the angle $\theta_{23}$ is generated without breaking the 
$Z_2$ symmetries and can be much smaller compared to $\theta_{12}$. 
The remaining angle $\theta_{13}$ is given by 
breaking the $Z_2$ symmetries with higher dimensional operators, which are fully specified 
within the considered model, providing an explanation for why 
it is more suppressed. In this way, the model provides a qualitative explanation for 
the smaller mixing angles, although their quantitative values must be  
fitted to experimental values, rather than being predicted.

This paper is organized as follows. In section~\ref{2} we discuss the 
symmetry $Z_n\times Z_m$ of the quark mass matrices and the relation with the CKM matrix.
In section~\ref{3} we review the group theory of the $\Delta(6N^2)$ series and identify 
suitable $Z_2 \times Z_2$ subgroups which may be preserved in the quark sector, leading to 
a successful determination of the Cabibbo angle. 
In section~\ref{4} we construct a model of quarks based on $\Delta(6N^2)$, the first of its kind
in the literature. We construct the quark mass matrices and resulting CKM mixing at the leading and 
next-to-leading order and derive the vacuum alignments that are required.
In section~\ref{5} we perform a full numerical analysis of the model for $N=28$ and show that all the 
quark masses and CKM parameters may be accommodated.
Section~\ref{6} summarises the paper.

\section{CKM matrix and $Z_n\times Z_m$ symmetry of quark mass matrices}
\label{2}
The quark mass matrices are defined in a general RL basis by
\begin{align}
-{\cal L}=
\begin{pmatrix}
\overline{u} & \overline{c} & \overline{t}
\end{pmatrix}_R
M_u 
\begin{pmatrix}
u\\
c\\
t\\
\end{pmatrix}_L
 + 
 \begin{pmatrix}
\overline{d} & \overline{s} & \overline{b}
\end{pmatrix}_R
M_d
\begin{pmatrix}
d\\
s\\
b\\
\end{pmatrix}_L + H.c..
\end{align}
We write the mass matrices in the diagonal basis with hats, where,
\begin{equation}
M_u = V_u' {\hat M}_u V_u^\dagger  \ \ {\rm and} \ \
 M_d = V_d' {\hat M}_d V_d^\dagger .
\end{equation}
Hence,
\begin{equation}
M_u^\dagger M_u = V_u {\hat M}_u^\dagger {\hat M}_u V_u^\dagger  \ \ {\rm and} \ \
 M_d^\dagger M_d = V_d {\hat M}_d^\dagger {\hat M}_d V_d^\dagger .
\end{equation}
In the diagonal basis the mass matrices are invariant under ${\hat Q}$ and ${\hat A}$ transformations, 
\begin{equation}
{\hat Q}^\dagger  \left({\hat M_u}^\dagger  {\hat M_u} \right) {\hat Q} = {\hat M_u}^\dagger  {\hat M_u} \ \ {\rm and} \ \
{\hat A}^\dagger  \left({\hat M_d}^\dagger  {\hat M_d} \right) {\hat A} = {\hat M_d}^\dagger  {\hat M_d} ,
\end{equation}
where ${\hat Q}$ and ${\hat A}$ are elements of $Z_n$ and $Z_m$, respectively, given by
\begin{eqnarray}
{\hat Q}=\begin{pmatrix}
e^{2\pi i n_u/n}&0&0\\
0&e^{2\pi i n_c/n}&0\\
0&0&e^{2\pi i n_t/n}\\
\end{pmatrix},
\quad
{\hat A}=\begin{pmatrix}
e^{2\pi i m_d/m}&0&0\\
0&e^{2\pi i m_s/m}&0\\
0&0&e^{2\pi i m_b/m}\\
\end{pmatrix},
\end{eqnarray}
where $n_{u,c,t}$ and $m_{d,s,b}$ are integers. 
It then follows that in the original (non-diagonal) basis that the mass matrices are invariant under $Q$ and 
$A$ transformations,
\begin{equation}
Q^\dagger  \left(M_u^\dagger  M_u \right) Q = M_u^\dagger  M_u \ \ {\rm and} \ \
A^\dagger  \left( M_d^\dagger  M_d \right) A = M_d^\dagger  M_d ,
\end{equation}
where 
\begin{eqnarray}
Q=V_u {\hat Q}
V_u^\dagger,
\quad
A=V_d {\hat A}
V_d^\dagger .
\end{eqnarray}
In the non-diagonal basis they also satisfy $Q^n=A^m= e$. Since the CKM matrix is given by $V_u^\dagger V_d$, it can be determined
from the matrices which diagonalise $Q$ and $A$,
\begin{eqnarray}
Q=V_Q {\hat Q}
V_Q^\dagger,
\quad
A=V_A {\hat A}
V_A^\dagger ,
\end{eqnarray}
where we identify $V_u=V_Q$ and $V_d=V_A$.

\section{The group $\Delta(6N^2)$ and $Z_2$ symmetry}
\label{3}

Let us shortly review the discrete group $\Delta (6N^2)$, which is  isomorphic 
to $(Z_N \times Z_N')\rtimes S_3$ \cite{Escobar:2008vc}.
We denote $S_3$ generators by $a$ and $b$, where
where $a$ and $b$ are $Z_3$ and $Z_2$, 
and the generators of $Z_N$ and $Z_N'$ by $a$ and $a'$. 
These generators  satisfy
\begin{eqnarray}
& & a^3=b^2=(ab)^2 =c^N = {d}^N =  e, 
\quad cd = dc, 
\nonumber \\
& &  aca^{-1}=c^{-1}d^{-1}, \quad ada^{-1}=c,
\nonumber \\
& &  bcb^{-1}=d^{-1}, \quad bdb^{-1}=c^{-1}.
\end{eqnarray}
Using them, all of $\Delta (6N^2)$ elements are written as 
\begin{eqnarray}
 & & g=a^{k}b^{\ell}c^{m}d^{n},
\end{eqnarray}
for $k=0,1,2$, $\ell=0,1$ and $m,n=0,1,2,\cdots ,N-1$.
The character table is written in Tab. \ref{tab:Delta-3N-character}.

\begin{table}[t]
\begin{center}
{\footnotesize
\begin{tabular}{|c|c|c|c|c|c|c|c|c|}
\hline
&$h$&$\chi_{1_r}$&$\chi_{2}$&$\chi_{3_{1k}}$&$\chi_{3_{2k}}$&$\chi_{6_{[[k],[\ell]]}}$\\
\hline
$C_1$&1&1 & 2&3&3&6 \\ 
\hline
$C_3^{(m)}$&$\frac{N}{\gcd(N,m)}$&1 & 2
&$\eta^{-2m k}+2\eta^{m k}$&$\eta^{-2m k}+2\eta^{m k}$ 
&$2\eta^{m (k-\ell)}+2\eta^{-m (2k+\ell)}+2\eta^{m(k+2\ell)}$ \\ 
\hline
$C_{6}^{(m,n)}$&$\frac{N}{\gcd(N,m,n)}$&1 &2 
&$\eta^{m k}+\eta^{-n k}$
&$\eta^{m k}+\eta^{-n k}$
&$\eta^{m k+n\ell}+\eta^{(-m+n)k-m\ell}+\eta^{-n k+(m-n)\ell}$  
\\ 
&& & 
&$+\eta^{(-m +n)k}$
&$+\eta^{(-m +n)k}$
&$+\eta^{-n k-m\ell}+\eta^{m k+(-m+n)\ell}+\eta^{(-m+n)k+n\ell}$  \\ 
\hline
$C_{2N^2}$&$3$&1 & $-1$&$0$     &0&0 \\ 
\hline
$C_{3N}^{(m)}$ &$\frac{2N}{\gcd(N,m)}$& $(-1)^r$ &$0$ 
&$\eta^{-m k}$&$-\eta^{-m k}$&0\\
\hline
\end{tabular}
}
\end{center}
\caption{Character table of $\Delta(6N^2)$ for $N/3 \neq $ integer, 
where $\eta=e^{2\pi i/N}$. }
\label{tab:Delta-3N-character}
\end{table}

For $N\not=$integer, irreducible representations are 
${\bf 1}_{0,1}$, $\bf 2$, ${\bf 3}_{1k}$, 
${\bf 3}_{2k}$, and ${\bf 6}_{[[k],[\ell]]}$. 
Tensor products relating to doublet and triplets are 
\begin{eqnarray}
\begin{split}
{\bf 3}_{1k}\times {\bf 3}_{1k'}
={\bf 3}_{1(k+k')}+{\bf 6}_{[[k],[-k']]},
\quad
{\bf 3}_{1k}\times {\bf 3}_{2k'}
={\bf 3}_{2(k+k')}+{\bf 6}_{[[k],[-k']]},
\\
{\bf 3}_{2k}\times {\bf 3}_{2k'}
={\bf 3}_{1(k+k')}+{\bf 6}_{[[k],[-k']]},
\quad
{\bf 3}_{1k}\times {\bf 2}
={\bf 3}_{1 k }+{\bf 3}_{2k},
\\
{\bf 3}_{2k}\times {\bf 2}
={\bf 3}_{1 k }+{\bf 3}_{2k},
\quad
{\bf 2}\times {\bf 2} 
={\bf 1}_{0}+{\bf 1}_{1}+{\bf 2}.
\end{split}
\end{eqnarray}
Some triplets and sextet are reducible, 
precisely ${\bf 3}_{10}={\bf 1}_0+{\bf 2}$, 
${\bf 3}_{20}={\bf 1}_1+{\bf 2}$, 
and ${\bf 6}_{[[-k],[k]]}={\bf 3}_{1k}+{\bf 3}_{2k}$. 
If their representations are explicitly given, they are 
$(x_1,x_2,x_3)_{{\bf 3}_{10}}=(x_1+x_2+x_3)_{{\bf 1}_0}
+(\omega x_1+x_2+\omega^2 x_3,\omega^2 x_1+x_2+\omega x_3)_{\bf 2}$, 
$(x_1,x_2,x_3)_{{\bf 3}_{20}}=(x_1+x_2+x_3)_{{\bf 1}_1}
+(\omega x_1+x_2+\omega^2 x_3,\omega^2 x_1+x_2+\omega x_3)_{\bf 2}$, 
and $(x_1,x_2,x_3,x_4,x_5,x_6)_{{\bf 6}_{[[-k],[k]]}}
=(x_1+x_6,x_2+x_5,x_3+x_4)_{{\bf 3}_{1k}}
+(-x_1+x_6,-x_2+x_5,-x_3+x_4)_{{\bf 3}_{2k}}$.

As residual symmetry, we will choose $Z_2$ symmetry 
in this group. The elements of $Z_2$ symmetry 
is belonging to the conjugacy class of 
\begin{eqnarray}
\begin{array}{cc}
     C_{3N}^{(\ell)}:&
     \left \{
     bc^{\ell+n}d^{n}, ~
     a^2bc^{-\ell}d^{-\ell-n},~
     abc^{-n}d^{\ell}
~     |
     ~n=0,1,\cdots,N-1
     \right \}, 
\\
&
~\ell=0,1,\cdots,N-1.
\end{array}
\end{eqnarray}
The number of this class is $3N$ for each 
choice of $\ell$ so that total number is $3N^2$. 
By taking $3\times 3$ matrix representations, 
the meaning of $3N^2$ is explained as follows.  
The three choices mean the choice of three angles 
$\theta_{12}$, $\theta_{13}$, and $\theta_{23}$ 
to be maximal mixing with some phase factor. 
The one of $N$ choices for the charge of $Z_N$ 
which determines the phase of maximal mixing. 
The last $N$ choices exist to determine the 
phase of trace. 

In matrix representation, the generators are written by
\begin{eqnarray}
a=\begin{pmatrix}
0&1&0\\
0&0&1\\
1&0&0\\
\end{pmatrix},
\quad
b=\pm \begin{pmatrix}
0&0&1\\
0&1&0\\
1&0&0\\
\end{pmatrix},
\quad
c=\begin{pmatrix}
\eta^k &0&0\\
0&\eta^{-k}&0\\
0&0&1\\
\end{pmatrix},
\quad
d=\begin{pmatrix}
1&0&0\\
0&\eta^k &0\\
0&0&\eta^{-k}\\
\end{pmatrix},
\quad
\end{eqnarray}
for the triplet ${\bm 3}_{1k}$ with plus sign 
and for ${\bm 3}_{2k}$ with minus sign
where $\eta=e^{2\pi i/N}$, 
Let us take specific choice for the symmetries 
of mass matrices $Q=abc^{x}$ and $A=abc^{y}$, i.e.
\begin{align}
Q=
\begin{pmatrix}
0&\eta^{-lx}&0\\
\eta^{lx}&0&0\\
0&0&1
\end{pmatrix},
\quad
A=
\begin{pmatrix}
0&\eta^{-ly}&0\\
\eta^{ly}&0&0\\
0&0&1
\end{pmatrix},
\end{align}
for ${\bm 3}_{1k}$ to $Q$ and 
${\bm 3}_{1l}$ to $A$. 
This specific choice makes $\theta_{12}$ to be maximal, 
the charge of $Z_N$ fixed, and the trace being $1$. 
Because of the degeneracy of eigenvalue for the above matrices,
we generally have
\begin{align}
Q=
V_Q
\begin{pmatrix}
-1&0&0\\
0&1&0\\
0&0&1
\end{pmatrix}
V_Q^\dagger,
\quad
A=
V_A
\begin{pmatrix}
-1&0&0\\
0&1&0\\
0&0&1
\end{pmatrix}
V_A^\dagger,
\end{align}
where 
\begin{eqnarray}
\begin{split}
V_Q
=\frac{1}{\sqrt2}\begin{pmatrix}
-\eta^{kx}&\eta^{kx}&0\\
1&1&0\\
0&0&\sqrt2
\end{pmatrix}
\begin{pmatrix}
1&0&0\\
0&\cos\theta&\sin\theta\\
0&-\sin\theta&\cos\theta
\end{pmatrix},
\\
V_A
=\frac{1}{\sqrt2}
\begin{pmatrix}
-\eta^{ly} &\eta^{ly}&0\\
1&1&0\\
0&0&\sqrt2
\end{pmatrix}
\begin{pmatrix}
1&0&0\\
0&\cos\theta'&\sin\theta'\\
0&-\sin\theta'&\cos\theta'
\end{pmatrix}.
\end{split}
\end{eqnarray}
As discussed in the previous section, 
the CKM matrix is given by 
$V_\text{CKM}=V_Q^\dagger V_A$ so that
\begin{eqnarray}
V_\text{ CKM}=
\frac12
\begin{pmatrix}
\eta^{-kx+ly}+c&-\eta^{-kx+ly}+c&\sqrt2 s\\
-\eta^{-kx+ly}+c &\eta^{-kx+ly}+c&\sqrt 2s \\
-\sqrt2 s&-\sqrt2 s&2c
\end{pmatrix},
\end{eqnarray}
where 
$c=\cos(-\theta+\theta')$, 
$s=\sin(-\theta+\theta')$ are undetermined. 
For simplicity, if we take $\theta=\theta'$, 
it predicts the Cabibbo angle as $\sin\theta_{C}=\sin(\pi(-kx+l y)/N)$. 
By choosing $N=14$ and $-kx+l y=1$, 
it is close to the best fit value $\theta_C\approx 0.22$. 
As a general problem for the model 
which preserves $Z_2$ symmetry, if the residual symmetry is unbroken, then 
$|V_{ub}|$ and $|V_{cb}|$ will have the same value, undetermined by symmetry.

In the work \cite{King:2013vna}, the lepton mixing is predicted 
by model independent method with $\Delta(6N^2)$. 
According to this, $\sin\theta_{13}=\sqrt{2/3}\sin(\pi \gamma'/N)$ or 
$\sin\theta_{13}=\sqrt{2/3}\cos(\pi/6\pm \pi \gamma'/N)$
where $\gamma'=1,\cdots,N/2$, $\theta_{23}=45^\circ\mp \theta_{13}/\sqrt2$. 
As it predicts tri-maximal mixing so that $\sin^2\theta_{12}\approx1/3$. 
Experimentally, the best fit value is close to $\sin\theta_{13}\approx 0.15$. 
Some values predicted by $N=14$  are 
$|\sin\theta_{13}|=0.122, 0.182$. 
In the case $N=28$, it can be closer to the experimental value
$|\sin\theta_{13}|=0.152$.

\section{The Model}
\label{4}
\subsection{Quark masses and mixing}
Assuming $N/3$ is not integer, 
the model we consider is defined in Tab.~\ref{table2}. 
\begin{table}[h]
\begin{tabular}{|c|ccccccccccccc|}
\hline
&$(q_1,q_2,q_3)$&$ u^c$&$c^c$&$t^c$
&$ d^c$&$s^c$&$b^c$ &$h_u, h_d$ &$\chi_u$  &$\chi_d$ &$\chi_3$
&$\chi_1$&$\chi_1'$\\ 
\hline
$\Delta(6N^2)$ & ${\bf 3}_{2k}$& ${\bf 1}_0$& ${\bf 1}_0$& ${\bf 1}_1$
& ${\bf 1}_0$& ${\bf 1}_0$&${\bf 1}_0$& ${\bf 1}_{0}$
&  ${\bf 3}_{1(-k)}$& ${\bf 3}_{1(-k)}$&  ${\bf 3}_{2(-2k)}$
& ${\bf 1}_{1}$& ${\bf 1}_{1}$
\\
$Z_{N+1}$ & $0$& $-2$& $-2$& $0$
& $-2$& $0$&$0$& $0$
& $1$&  $0$& $0$
& $2$& $0$
\\
$Z_{N+1}$ & $0$& $-2$& $0$& $0$
& $-2$& $-2$&$-2$& $0$
& $0$&  $1$& $0$
& $0$& $2$
\\
$U(1)_R$ & $1$& $1$& $1$& $1$
& $1$& $1$&$1$& $0$
& $0$&  $0$& $0$
& $0$& $0$
\\
\hline
\end{tabular}
\caption{Charge assignment of quarks, Higgs, and flavors for the flavor symmetry 
$\Delta(6N^2)\times Z_{N+1}\times Z_{N+1}$ and $U(1)_R$.}
\label{table2}
\end{table}

We take vacuum expectation values for all the scalar fields 
and assume vacuum alignment such that
\begin{eqnarray}
\begin{split}
\langle h_u\rangle=v_u,
\quad
\langle h_d\rangle=v_d,
\quad
\langle \chi_1\rangle=u_4,
\quad
\langle \chi_1'\rangle=u_5,
\\
\langle\chi_u\rangle
=\begin{pmatrix}
u_1\\
u_1 \eta^{-kx}\\
0
\end{pmatrix},
\quad
\langle\chi_d\rangle
=\begin{pmatrix}
u_2\\
u_2 \eta^{-ky}\\
0
\end{pmatrix},
\quad
\langle\chi_3\rangle
=\begin{pmatrix}
0\\
0\\
u_3
\end{pmatrix}.
\end{split}
\end{eqnarray}
The residual symmetries are 
$Q=abc^x$ for up-type quarks 
and $A=abc^y$ for down-type quarks. 
Considering the triplet ${\bf 3}_{1(-k)}$ representation, 
we have
\begin{eqnarray}
Q=
\begin{pmatrix}
0&\eta^{kx}&0\\
\eta^{-kx}&0&0\\
0&0&1\\
\end{pmatrix},
\quad
A
=\begin{pmatrix}
0&\eta^{ky}&0\\
\eta^{-ky}&0&0\\
0&0&1\\
\end{pmatrix}.
\end{eqnarray}
Then we have 
$Q\langle\chi_u\rangle=\langle\chi_u\rangle$, 
$A\langle\chi_u\rangle\not=\langle\chi_u\rangle$, 
and $A\langle\chi_d\rangle=\langle\chi_d\rangle$, 
$Q\langle\chi_d\rangle\not=\langle\chi_d\rangle$.

The allowed Yukawa couplings are
\begin{eqnarray}
\begin{split}
{\cal L}
=&\frac{y_{u1}}{\Lambda^2}
(q_1,q_2,q_3)
c^c h_u
\chi_u^2
+\frac{y_{u2}}{\Lambda^2}
(q_1,q_2,q_3)
c^c h_u
\chi_3\chi_1
+\frac{y_{u3}}{\Lambda}
(q_1,q_2,q_3)
t^c h_u
\chi_3
\\
&+\frac{y_{d1}}{\Lambda}
(q_1,q_2,q_3)
 s^c h_d
\chi_d^2
+\frac{y_{d2}}{\Lambda}
(q_1,q_2,q_3)
s^c h_d
\chi_3\chi_1'
+\frac{y_{d3}}{\Lambda}
(q_1,q_2,q_3)
b^c h_d
\chi_d^2
+\frac{y_{d4}}{\Lambda}
(q_1,q_2,q_3)
b^c h_d
\chi_3\chi_1',
\end{split}
\end{eqnarray}
where $\Lambda$ is the cutoff scale. 
The multiplication of ${\bf 3}_{1k}$ and  ${\bf 3}_{1(-k)}$
is $(x_1,x_2,x_3)_{{\bf 3}_{1k}}\times
(y_1,y_2,y_3)_{{\bf 3}_{1(-k)}}
=(x_1y_1+x_2y_2+x_3y_3)_{{\bf 1}_{0}}
+(\omega x_1y_1+x_2y_2+\omega^2x_3y_3,
\omega^2 x_1y_1+x_2y_2+\omega x_3y_3)_{{\bf 2}}$.

Then mass matrices become
\begin{eqnarray}
(M_u)_{RL}
=\frac{v_u}{\Lambda^2}\begin{pmatrix}
 0&0&  0\\
 y_{u1}u_1^2&y_{u1}u_1^2\eta^{-2kx}& y_{u2}u_3u_4\\
0&0&y_{u3}u_3\Lambda \\
\end{pmatrix},
\quad
(M_d)_{RL}
=\frac{v_d}{\Lambda^2}\begin{pmatrix}
 0&0& 0\\
 y_{d1}u_2^2&y_{d1}u_2^2\eta^{-2ky}& y_{d2}u_3u_5\\
 y_{d3}u_2^2&y_{d3}u_2^2\eta^{-2ky}& y_{d4}u_3u_5\\
\end{pmatrix}.
\end{eqnarray}
They are rank 2 matrices so 
one eigenvalue is vanishing for each sector.  
Assuming all the Yukawa couplings are real,  
mass matrices in $LL$ basis can be
diagonalised by $V_{u}=V_{23}^u V_{12}$ and 
$V_d=V_{23}^dV_{12}'$. 
Then, the CKM matrix has the form
\begin{eqnarray}
V_{CKM}
=\frac12
\begin{pmatrix}
1+c_{23} \eta^{2k(x-y)} &-c_{23}\eta^{2kx}+\eta^{2ky} &-\sqrt2 s_{23}\eta^{2kx} \\
\eta^{-2kx}-c_{23}\eta^{-2ky} &c_{23}+\eta^{2k(-x+y)} &\sqrt2 s_{23}\\
\sqrt2 s_{23}\eta^{-2ky}&-\sqrt2 s_{23}&2c_{23}\\
\end{pmatrix}.
\end{eqnarray}

\subsection{Next-to-next-to-leading correction}

Correction terms of higher dimensional operators are 
\begin{eqnarray}
\begin{split}
\Delta {\cal L}
=&\frac{y_{u4}}{\Lambda^3}
(q_1,q_2,q_3)
 u^c h_u
\chi_u\chi_1\chi_1'
+\frac{y_{u5}}{\Lambda^3}
(q_1,q_2,q_3)
 u^c h_u
\chi_d\chi_1^2
\\
&+\frac{y_{d5}}{\Lambda^3}
(q_1,q_2,q_3)
d^c h_d
\chi_u\chi_1'^2
+\frac{y_{d6}}{\Lambda^3}
(q_1,q_2,q_3)
d^c h_d
\chi_d\chi_1\chi_1'.
\end{split}
\end{eqnarray}

Then mass matrices become
\begin{eqnarray}
\begin{split}
(M_u)_{RL}
=\frac{v_u}{\Lambda^2}\begin{pmatrix}
 0&0&  0\\
 y_{u1}u_1^2&y_{u1}u_1^2\eta^{-2kx}& y_{u2}u_3u_4\\
0&0&y_{u3}u_3\Lambda \\
\end{pmatrix}
\\
+\frac{u_4v_u}{\Lambda^3}
\begin{pmatrix}
 y_{u4}u_1u_5+y_{u5}u_2u_4& y_{u4}u_1u_5\eta^{-kx}+y_{u5}u_2u_4\eta^{-ky}& 0\\
0&0& 0\\
0&0&0\\
\end{pmatrix},
\\
(M_d)_{RL}
=\frac{v_d}{\Lambda}\begin{pmatrix}
 0&0& 0\\
 y_{d1}u_2^2&y_{d1}u_2^2\eta^{-2ky}& y_{d2}u_3u_5\\
 y_{d3}u_2^2&y_{d3}u_2^2\eta^{-2ky}& y_{d4}u_3u_5\\
\end{pmatrix}
\\
+\frac{ u_5 v_d}{\Lambda^3}
\begin{pmatrix}
 y_{d5}u_1u_5+y_{d6}u_2u_4
 & y_{d5}u_1u_5\eta^{-kx}+y_{d6}u_2u_4\eta^{-ky}& 0\\
0&0& 0\\
0&0&0\\
\end{pmatrix}.
\end{split}
\end{eqnarray}
These are rank 3 matrices and break 
$Z_2$ symmetry, then we 
obtain up and down masses and 
$\theta_{13}$.

Now we have all the mixing angles from up and down quarks. 
Using $V_u=V_{12}V_{23}^uV_{13}^u$ and 
$V_d=V_{12}'V_{23}^dV_{13}^d$, 
the CKM matrix becomes 
$V_{\rm CKM}=V_{12}^\dagger V_{23}^{u\dagger}V_{13}
V_{23}^d V_{12}'$ where 
$V_{13}=V_{13}^{u\dagger}V_{13}^d$. 
To find out $\theta_{13}$ and $\theta_{23}$ for CKM matrix, 
let us consider
\begin{eqnarray}
\begin{split}
V_{ub}
&=\frac{1}{\sqrt2}
(-\sin\theta_{23}^d\cos\theta_{23}^u\eta^{kx}
+\cos\theta_{13}^{ud}\sin\theta_{23}^u\cos\theta_{23}^d\eta^{kx}
+\sin\theta_{13}^{ud}\cos\theta_{23}^d
),
\\
V_{cb}
&=\frac{1}{\sqrt2}
(\sin\theta_{23}^d\cos\theta_{23}^u
-\cos\theta_{13}^{ud}\sin\theta_{23}^{u}\cos\theta_{23}^d
+\sin\theta_{13}^{ud}\cos\theta_{23}^d\eta^{-kx}
),
\end{split}
\end{eqnarray}
where $\theta_{23}^u$ is the angle of $V_{23}^u$, 
$\theta_{23}^d$ is the angle of $V_{23}^d$, 
and $\theta_{13}^{ud}$ is the angle of $V_{13}^{ud}$. 
Assuming these mixing angles are small, 
$|V_{ub}|$ and $|V_{cb}|$ can be expanded by
\begin{eqnarray}
\begin{split}
|V_{ub}|
&=\frac{1}{\sqrt2}
\sqrt{(\theta_{23}^d-\theta_{23}^u)^2
+(\theta_{13}^{ud})^2
-2\theta_{13}^{ud}(\theta_{23}^d-\theta_{23}^u)
\cos(kx)}),
\\
|V_{ub}|
&=\frac{1}{\sqrt2}
\sqrt{(\theta_{23}^d-\theta_{23}^u)^2
+(\theta_{13}^{ud})^2
+2\theta_{13}^{ud}(\theta_{23}^d-\theta_{23}^u)
\cos(kx)}).
\end{split}
\end{eqnarray}
By tuning the angles, we will obtain $|V_{ub}|\ll |V_{cb}|$.

\subsection{Potential analysis with driving field}
\begin{table}[h]
\begin{tabular}{|c|cccccccccccc|}
\hline
 &$\chi_u$ \  &${ \chi}_d$ &${ \chi}_3$ 
 &$\chi_1$ \  &${ \chi}_1'$  
  &$\Phi_1$ &$\Phi_2$&$\Phi_3$&$\Phi_4$ 
 &$\Phi_5$ &$\Phi_6$&$\Phi_7$
 \\ 
\hline
$\Delta(6N^2)$ 
& ${\bf 3}_{1(-k)}$&  ${\bf 3}_{1(-k)}$&  ${\bf 3}_{2(-2k)}$
& ${\bf 1}_{1}$&  ${\bf 1}_{1}$&  ${\bf 1}_{1}$
& ${\bf 1}_{0}$&  ${\bf 1}_{0}$&  ${\bf 1}_{0}$
&  ${\bf 1}_{0}$ 
& ${\bf 1}_{0}$&  ${\bf 1}_{0}$
\\
$Z_{N+1}$ &$1$ & $0$
&$0$&$2$&$0$
& $0$&$-2$& $-3$
&$1$&$0$& $0$&$0$
\\
$Z_{N+1}$ &$0$& $1$
&$0$&$0$&$2$& $0$&$0$& $0$
&$0$&$-2$& $-3$&$1$
\\
$U(1)_R$ & $0$&$0$ &$0$ 
&$0$&$0$
& $2$&  $2$& $2$&  $2$
& $2$&  $2$&  $2$
\\
\hline
\end{tabular}
\caption{Charge assignment of flavors and driving fields for the flavor symmetry 
$\Delta(6N^2)\times Z_{N+1}\times Z_{N+1}$ and $U(1)_R$ symmetry.}
\end{table}

We introduce driving fields $\Phi_u$ and $\Phi_d$. 
The super potential becomes
\begin{eqnarray}
\begin{split}
w=\frac{\lambda_1}{\Lambda} \chi_3^3 \Phi_1
+\frac{\lambda_2}{\Lambda^2} \chi_u^2\chi_3^2 \Phi_2
+\frac{\lambda_3}{\Lambda^2} \chi_3^3\chi_1 \Phi_2
+\frac{\lambda_4}{\Lambda} \chi_u^3 \Phi_3
+\frac{\lambda_5}{\Lambda^{N-2}}
\sum_n( \chi_u^{N-12n} \chi_1^{12n})\Phi_4
\\
+\frac{\lambda_6}{\Lambda^2} \chi_d^2\chi_3^2 \Phi_5
+\frac{\lambda_7}{\Lambda^2} \chi_3^3\chi_1' \Phi_5
+\frac{\lambda_8}{\Lambda} \chi_d^3 \Phi_6
+\frac{\lambda_9}{\Lambda^{N-2}} 
\sum_n( \chi_d^{N-12n} \chi_1'^{12n}) \Phi_7.
\end{split}
\end{eqnarray}
They are explicitly written by
\begin{eqnarray}
\begin{split}
w=&
\frac{\lambda_1}{\Lambda}\chi_{31}\chi_{32}\chi_{33}\Phi_{1}
+\frac{\lambda_2}{\Lambda^2}
(\chi_{u1}^2\chi_{32}\chi_{33}+\chi_{u2}^2\chi_{33}\chi_{31}
+\chi_{u3}^2\chi_{31}\chi_{32})\Phi_{2}
+\frac{\lambda_3}{\Lambda^2}
\chi_{31}\chi_{32}\chi_{33}\chi_{1}\Phi_{2}
\\&
+\frac{\lambda_4}{\Lambda}\chi_{u1}\chi_{u2}\chi_{u3}\Phi_{3}
+\frac{\lambda_5}{\Lambda}
(\chi_{u1}^N+\chi_{u2}^N+\chi_{u3}^N
+\sum_n(\chi_{u1}\chi_{u2}\chi_{u3})^{N-12n}
\chi_1^{12n})\Phi_{4}
\\&
+\frac{\lambda_6}{\Lambda^2}
(\chi_{d1}^2\chi_{32}\chi_{33}+\chi_{d2}^2\chi_{33}\chi_{31}
+\chi_{d3}^2\chi_{31}\chi_{32})\Phi_{5}
+\frac{\lambda_7}{\Lambda^2}
\chi_{31}\chi_{32}\chi_{33}\chi_1'\Phi_{5}
\\&
+\frac{\lambda_8}{\Lambda}\chi_{d1}\chi_{d2}\chi_{d3}\Phi_{6}
+\frac{\lambda_9}{\Lambda}
(\chi_{d1}^N+\chi_{d2}^N+\chi_{d3}^N
+\sum_n(\chi_{d1}\chi_{d2}\chi_{d3})^{N-12n}
\chi_1'^{12n})\Phi_{7}
\end{split}
\end{eqnarray}
The potential minimum conditions are
\begin{eqnarray}
\begin{split}
&\chi_{31}\chi_{32} \chi_{33}
=0,
\quad
\chi_{u1}^2\chi_{32} \chi_{33}
+\chi_{u2}^2\chi_{33} \chi_{31}
+\chi_{u3}^2\chi_{31} \chi_{32}
=0,
\\
&\chi_{u1}\chi_{u2} \chi_{u3}
=0,
\quad
\chi_{u1}^N+\chi_{u2}^N+ \chi_{u3}^N
=0,
\\
&\chi_{d1}^2\chi_{32} \chi_{33}
+\chi_{d2}^2\chi_{33} \chi_{31}
+\chi_{d3}^2\chi_{31} \chi_{32}
=0,
\\
&\chi_{d1}\chi_{d2} \chi_{d3}
=0,
\quad
\chi_{d1}^N+\chi_{d2}^N+ \chi_{d3}^N
=0.
\end{split}
\end{eqnarray}
We take the vacuum expectation values as
$\langle \chi_u\rangle=(u_1,u_2,u_3)$,  
$\langle \chi_d\rangle=(u_4,u_5,u_6)$, and 
$\langle \chi_3\rangle=(u_7,u_8,u_9)$.
At first, we need to choose one of $u_1$, $u_2$, 
and $u_3$ is zero, and similarly one of $u_4$, $u_5$, 
and $u_6$ is zero. Let us take $u_3=u_6=0$, 
then remaining equations are 
\begin{eqnarray}
\begin{split}
u_7u_8u_9=0,
\quad
u_1^2u_8u_9
+u_2^2u_9u_7
=0,
\quad
u_1^N=-u_2^N,
\\
u_4^2u_8u_9
+u_5^2u_9u_7
=0,
\quad
u_4^N=-u_5^N,
\end{split}
\end{eqnarray}
There are twp choices to satisfy all of them, 
$u_9=0$ or $u_7=u_8=0$, 
and we take the latter case. 
Then the vacuum alignment  that satisfies the conditions is
\begin{eqnarray}
\begin{split}
\langle\chi_u\rangle
=\begin{pmatrix}
u_1\eta^x\\
-u_1\eta^{x'}\\
0\\
\end{pmatrix},
\quad
\langle\chi_d\rangle
=\begin{pmatrix}
u_4\eta^y\\
-u_4\eta^{y'}\\
0\\
\end{pmatrix},
\quad
\langle\chi_3\rangle
=\begin{pmatrix}
0\\
0\\
u_9\\
\end{pmatrix},
\end{split}
\end{eqnarray}
where $x$, $x'$, $y$, and $y'$ are any integers. 
Therefore we can take the vacuum alignment used in our model.

\section{Numerical results}
\label{5}

With the next-to-next-to-leading corrections, we have 
11 Yukawa couplings and two phase parameters. 
Taking $y_{u3}u_3$ and $y_{d4}u_3u_5/\Lambda$ as common 
factors which can be fitted by top and bottom masses, 
we have 9 parameters. 
Precisely, the parameters for next leading order corrections 
for up quarks are $y_{u1}u_1^2/y_{u3}u_3\Lambda$  
and $y_{u2}u_4/y_{u3}\Lambda$.
For down quarks they are
$y_{d1}u_2^2/y_{d4}u_3u_5$,
$y_{d2}/y_{d4}$,
and $y_{d3}u_2^2/y_{d4}u_3u_5$.
NNLO corrections for up quarks are
$y_{u4}u_1u_5/y_{u3}u_3\Lambda$,
and $y_{u5}u_2u_4/y_{u3}u_3\Lambda$. 
NNLO corrections for down quarks are
$y_{d5}u_1u_5/y_{d4}u_3u_5$,
and $y_{d6}u_2u_4/y_{d4}u_3u_5$.
For the phases, we choose $N=28$ and 
$k(x-y)=2$ then we predict 
$\sin\theta_{12}=0.222521$ at the leading order.

We derive physical values, masses and mixing at the GUT scale. 
After renormalization group running, 
following values will be preferred by experiments \cite{Antusch:2013jca}:
\begin{eqnarray}
\begin{split}
\label{eqreg}
\theta_{12}\approx 0.2276,
\quad
2.9\times 10^{-3}\leq\theta_{13}\leq
3.4\times 10^{-3},
\quad
3.3\times 10^{-2}\leq\theta_{23}\leq
3.9\times 10^{-2},
\\
4.8\times 10^{-6}\leq\frac{m_u}{m_t}
\leq 5.4\times 10^{-6},
\quad
2.3\times 10^{-3}\leq\frac{m_c}{m_t}
\leq2.6\times 10^{-3},
\\
6.3\times10^{-4}\leq\frac{m_d}{m_b}
\leq 8.9\times10^{-4},
\quad
1.8\times10^{-2}\leq\frac{m_s}{m_b}
\leq 1.2\times10^{-2},
\end{split}
\end{eqnarray}
where we have chosen $1\leq\tan\beta\leq 50$, 
$-0.2\leq\bar\eta_{b},\bar\eta_{q}\leq 0.2$.

\begin{figure}
\includegraphics[width=8cm]{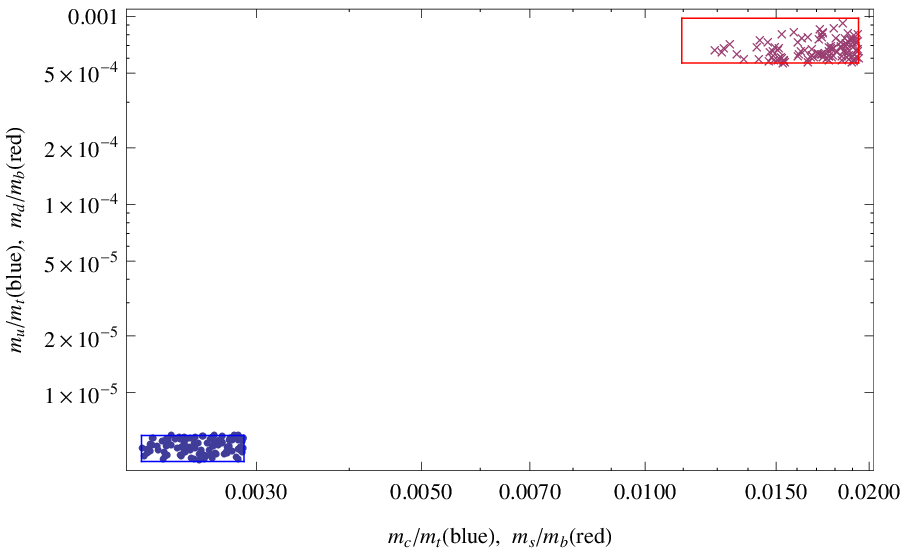}
\quad
\includegraphics[width=8cm]{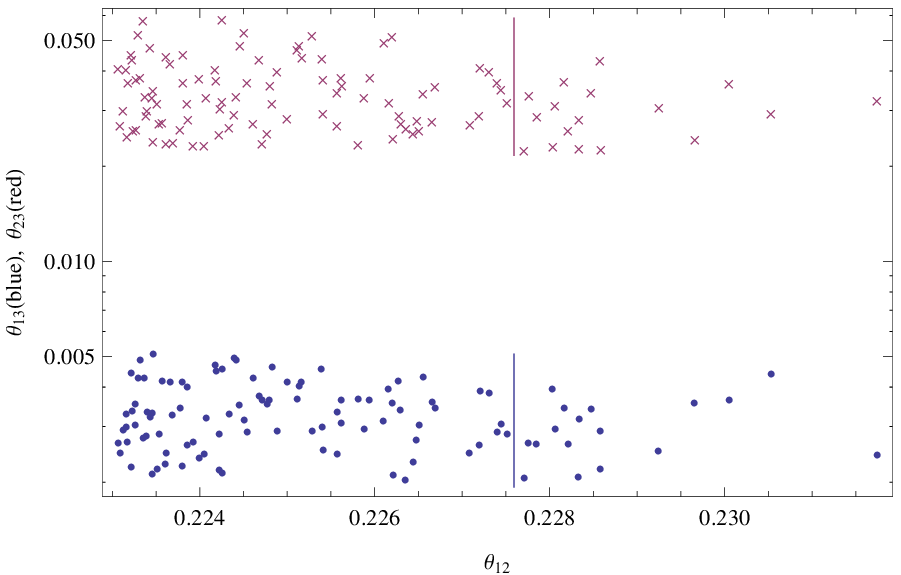}
\\ \\
\includegraphics[width=8cm]{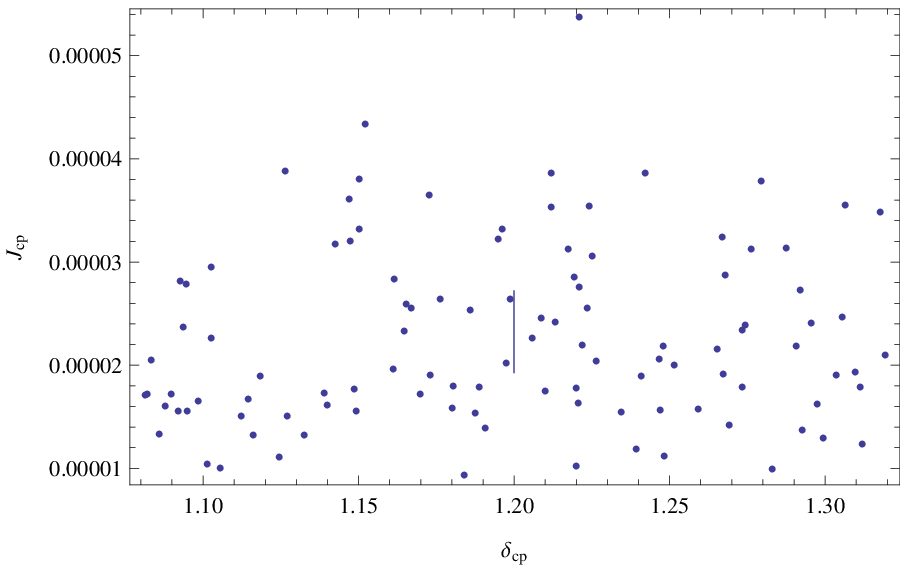}
\caption{Relations of physical parameters 
at the GUT scale and they will be allowed by experiments 
after running to the electroweak scale. 
The lines denote the allowed region including the threshold corrections 
$-0.2<{\bar\eta}_b,{\bar\eta}_q<0.2$ \cite{Antusch:2013jca}. 
For $\theta_{12}$ and $\delta_{CP}$, the running effects are 
small so we take around the best fit values.}
\end{figure}

\begin{figure}
\includegraphics[width=8cm]{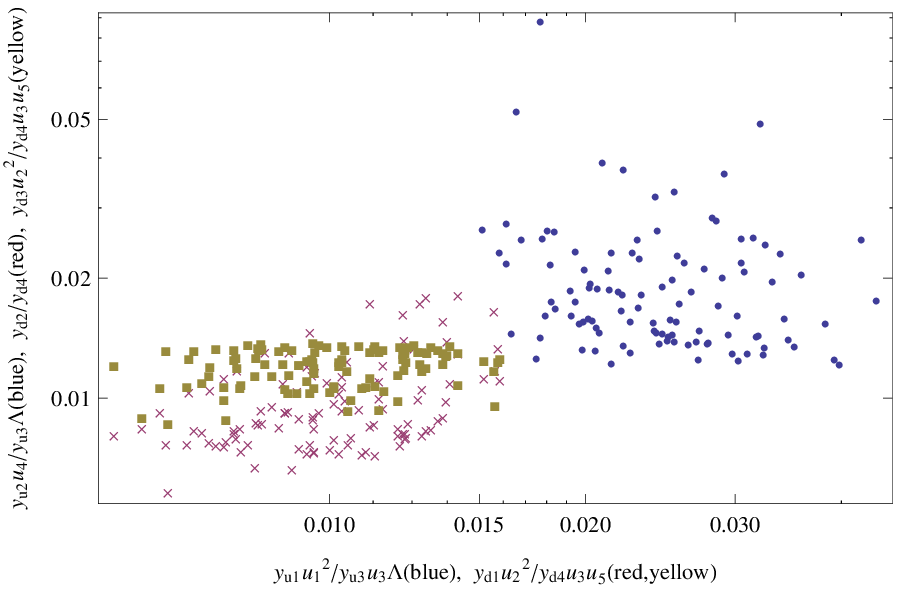}
\quad
\includegraphics[width=8cm]{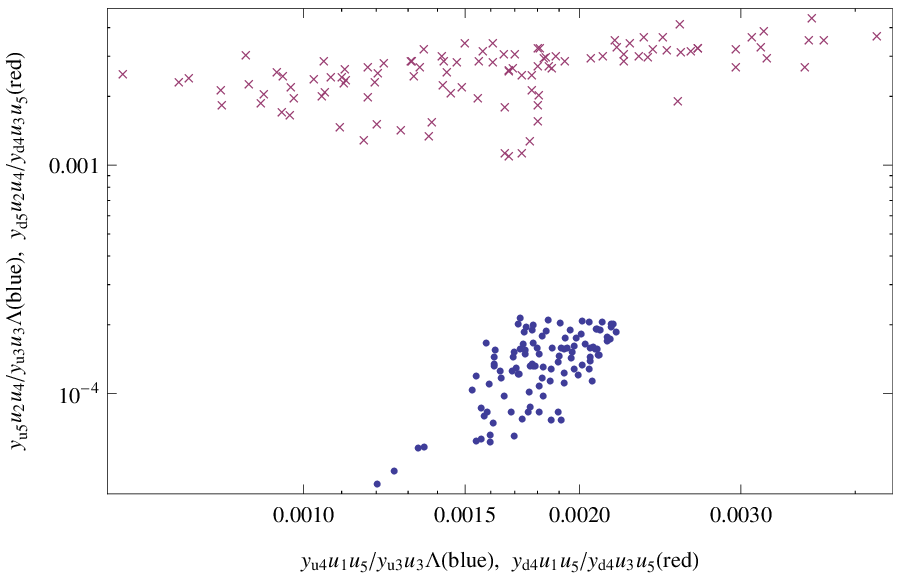}
\caption{Input values that satisfy the experimentally allowed region 
indicated by Fig. 1 are displayed. 
The number of Yukawa couplings is eleven and 
two parameters are chosen to be common factors 
to be fixed by the masses of top and bottom. 
Giving the rest nine parameters with the ratios of the two 
common factors, we get the physical values.
All the Yukawa couplings are considered to be complex 
and figures show absolute values.}
\end{figure}

In the figures 1 and 2, we show the random plots. 
Giving random values for all the Yukawa couplings and 
VEVs of flavons, we get physical values for masses and mixing by 
diagonalising mass matrices of up- and down-type quarks. 
We constrain the results to be consistent with experimental 
values indicating from Eq. (\ref{eqreg}). 
The physical values are actually three up-quark masses, three-down quark masses, 
three mixing angles, and CP phase. 
Since the third generation masses can be determined independently, we take mass ratios. 
For the convenience of numerical calculation, 
it includes 2$\%$ error for $\theta_{12}$ and 10$\%$ error for $\delta_{\rm CP}$. 
Expecting higher order corrections, these parameters will have some deviations 
and the errors will be reasonable. For Jarlskog invariant, we take no constraint and it 
is calculated by other parameters.

Fig. 2 show the parameter region of all the parameters we use 
for the mass matrices and all the points satisfy the constraints of Fig. 1. 
Since the Yukawa couplings are 
always appeared as the combinations with some flavon VEVs 
so we take ratios for the parameters with two chosen 
common factors $y_{u3}u_3\Lambda$ for up quarks 
and $y_{d4}u_3u_5$ for down quarks.
These two parameters can be given by fitting 
the third generation masses, top and bottom.
The left figure indicates NLO corrections which are of order $10^{-2}$ 
and the right figure is for NNLO corrections which are of order $10^{-3}$. 
The perturbation for the model seems successful.

\section{Summary}
\label{6}
We have proposed the first model of quarks in the literature 
based on the discrete
family symmetry $\Delta (6N^2)$
in which the Cabibbo angle is correctly determined
by a residual $Z_2\times Z_2$ subgroup,
and the smaller quark mixing angles may be qualitatively understood from the details of the model.
We emphasise that a concrete model
is required in order to shed light on the remaining small quark mixing angles
$\theta_{23}$ and $\theta_{13}$ which are not fixed by the symmetry alone.
In the present model we have performed a full numerical analysis for $N=28$ which shows
that all the quark masses and CKM parameters may be accommodated.

Unlike the dihedral groups, $\Delta (6N^2)$ contains triplet representations
and is capable of fixing all the lepton mixing angles using the direct approach. 
The present model of quarks may therefore be regarded as a first 
step towards formulating a complete model of quarks and leptons based on $\Delta (6N^2)$,
in which the lepton mixing matrix is fully determined by a Klein subgroup.
Taking $N=28$, such a model is capable of predicting $\sin\theta^{\rm MNS}_{13}=0.152$, 
$\sin\theta^{\rm CKM}_{12}=0.223$ at the leading order. 
As a general strategy, one can take any value for $\sin\theta^{\rm CKM}_{23}$ 
without breaking $Z_2$ symmetry and the smallest angle $\sin\theta^{\rm CKM}_{13}$ can be 
derived by NNLO terms which break $Z_2$.

\bigskip

\section*{Acknowledgement}

This work was supported in part by the Grant-in-Aid for Scientific Research 
No. 23.696 (H.I.). SFK acknowledges support from the 
European Union FP7 ITN-INVISIBLES (Marie Curie Actions, PITN- GA-2011- 289442) and the STFC Consolidated ST/J000396/1 grant.

\end{document}